\documentclass[aps,twocolumn,preprintnumbers,amsmath,amssymb,superscriptaddress,floatfix,nofootinbib]{revtex4}
\usepackage{graphicx,color,dcolumn,booktabs,bm}
\usepackage{longtable,lscape}
\usepackage{txfonts}
\usepackage{overpic}
\usepackage{epsfig}
\usepackage{amssymb}
\usepackage{rotating}
\usepackage{epstopdf}
\usepackage{appendix}
\usepackage{indentfirst}
\usepackage{feynmf}   
\usepackage{slashed}  
\usepackage{cases}
\usepackage{color}
\usepackage{multirow}
\usepackage{graphicx,color,dcolumn,booktabs,bm}
\usepackage{cases}
\usepackage{array}

\graphicspath{{Figures/}} %

\usepackage[colorlinks, citecolor=blue,anchorcolor=red,menucolor=red, linkcolor=red,filecolor=red,runcolor=red,urlcolor=blue,frenchlinks=red]{hyperref}

\begin{document}
\title{Understanding the nature of $\Omega(2012)$ in a coupled-channel approach}
\author{Qi-Fang L\"u} \email{lvqifang@hunnu.edu.cn} %
\affiliation{ Department of Physics, Hunan Normal University, and Key Laboratory of Low-Dimensional Quantum Structures and Quantum Control of Ministry of Education, Changsha 410081, China }

\affiliation{Key Laboratory for Matter Microstructure and Function of Hunan Province, Hunan Normal University, Changsha 410081, China}

\affiliation{Research Center for Nuclear Physics (RCNP), Ibaraki, Osaka 567-0047, Japan}

\author{Hideko Nagahiro} \email{nagahiro@rcnp.osaka-u.ac.jp} %

\affiliation{Department of Physics, Nara Women’s University, Nara 630-8506, Japan}

\affiliation{Research Center for Nuclear Physics (RCNP), Ibaraki, Osaka 567-0047, Japan}

\author{Atsushi Hosaka} \email{hosaka@rcnp.osaka-u.ac.jp} %

\affiliation{Research Center for Nuclear Physics (RCNP), Ibaraki, Osaka 567-0047, Japan}

\affiliation{Advanced Science Research Center, Japan Atomic Energy Agency, Tokai, Ibaraki 319-1195, Japan}

\begin{abstract}
We perform a systematic analysis of the mass and strong decays of the $\Omega(2012)$ resonance 
in a coupled-channel approach of a bare three-quark state and a composite meson-baryon state. 
Besides the coupling of the bare state and $\Xi^* \bar K$ channel, the effective meson-baryon interactions are also included. 
We find a pole with mass of 
$2008$ MeV 
by solving the Bethe-Salpeter equation. 
The calculated decay widths and its ratios of branching fractions $\mathcal {R}_{\Xi \bar K}^{\Xi \pi \bar K}$ agree well 
with the new experimental measurement by Belle Collaboration. 
Our results suggest that both three-quark core and $\Xi^* \bar K$ channel are essential for the description of the $\Omega(2012)$ resonance. 
We expect that present calculations can help us to better understand the nature of  $\Omega(2012)$ resonance.   

\end{abstract}

\maketitle

\section{Introduction}

Searching for more missing baryons and understanding their spectra and structures is an intriguing topics in hadron physics. 
In recent years, significant progresses have been achieved in the baryon spectroscopy both experimentally and theoretically, especially in the heavy quark sectors. 
Under this circumstance, the study on multistrangeness baryons is also experiencing a renaissance owing to the discoveries of new $\Xi$ and $\Omega$ resonances by the Belle, BESIII, and LHCb Collaborations~\cite{BESIII:2015dvj,Belle:2018mqs,Belle:2018lws,BESIII:2019cuv,LHCb:2020jpq}. 
The spectroscopy of multistrangeness baryons actually provides a connecting bridge between light and heavy baryons, which helps us to understand the variation due to different quark masses more deeply. Even so, our knowledge about these multistrangeness baryons is still far from complete as compared with the other baryons~\cite{ParticleDataGroup:2022pth}. More experimental information in this energy region are urgently needed, and it is also necessary to carefully study the properties of  every observed multistrangeness baryon so far.

In 2018, the Belle Collaboration observed a new structure $\Omega(2012)$ in the $\Xi^0 K^-$ and $\Xi^- K^0_S$ invariant mass distributions~\cite{Belle:2018mqs}. 
Subsequently, the Belle Collaboration has analyzed  the three-body decay of the $\Omega(2012)$ baryon to $\bar K \pi \Xi$, and give a $90\%$ credibility level upper limits on the ratios of the branching fractions relative to $\bar K \Xi$ decay modes~\cite{Belle:2019zco}. 
Also, the evidence of $\Omega(2012)$ in the $\Omega_c$ weak decay process was reported by Belle Collaboration in 2021~\cite{Belle:2021gtf}. Recently, with the $\Upsilon(1S)$,  $\Upsilon(2S)$, and $\Upsilon(3S)$ data collected by the Belle detector, the branching fraction ratio of the three-body decay to the two-body decay is measured~\cite{Belle:2022mrg}. 
Until now, the observed mass and decay widths of $\Omega(2012)$ are 
\begin{eqnarray}
	M [\Omega(2012)] = 2012.5 \pm 0.7 \pm 0.5~\rm{MeV},
\end{eqnarray} 
\begin{eqnarray}
  \Gamma[\Omega(2012)] = 6.4^{+2.5}_{-2.0}(stat) \pm 1.6(syst) ~\rm{MeV}.
\end{eqnarray} 
Defining the ratio of the branching fractions
\begin{eqnarray}
	\mathcal {R}_{\Xi \bar K}^{\Xi \pi \bar K} \equiv  \frac{\mathcal{B}[\Omega(2012) \to \Xi(1530) \bar K \to \Xi \pi \bar K]}{\mathcal{B}[\Omega(2012) \to \Xi \bar K]},
\end{eqnarray} 
one has the experimental data $\mathcal {R}_{\Xi \bar K}^{\Xi \pi \bar K}  ~\textless~11.9\% $~\cite{Belle:2019zco} in 2019 and $\mathcal {R}_{\Xi \bar K}^{\Xi \pi \bar K} = 0.97 \pm 0.24 \pm 0.07 $~\cite{Belle:2022mrg} in 2022. 
In the following, we employ $\Xi^*$ to denote the $\Xi(1530)$ by convention.

Before the discovery of the $\Omega(2012)$ resonance, there were already some predictions 
for the mass spectrum of the $\Omega$ family~\cite{Chao:1980em,Kalman:1982ut,Capstick:1986bm,Carlson:2000zr,Bijker:2000gq,Loring:2001ky,Schat:2001xr,Goity:2003ab,Matagne:2006zf,Oh:2007cr,Pervin:2007wa,Chen:2009de,An:2013zoa,Engel:2013ig,An:2014lga,Faustov:2015eba,Bijker:2015gyk,Liang:2015bxr,Aliev:2016jnp}, which suggested that the first orbital $P-$wave excitation should lie around 2 GeV. 
Then, the observations of the new multistrangeness resonance $\Omega(2012)$ has immediately interested a series of theoretical investigations. 
Among these works, two interpretations exist.
One is the conventional three-quark state with $P-$wave excitation with $J^P=3/2^-$~\cite{Xiao:2018pwe,Wang:2018hmi,Aliev:2018yjo,Aliev:2018syi,Polyakov:2018mow,Liu:2019wdr,Menapara:2021vug,Arifi:2022ntc,Wang:2022zja,Zhong:2022cjx}, which is denoted as $\Omega(1^2P_{3/2^-})$ following the standard notation $n ^{2S+1}\!L_J$ of spectroscopy.
The other is a meson-baryon molecular picture mainly contributed by $\Xi^* \bar K$ because of the proximity of their masses~\cite{Pavao:2018xub,Lin:2018nqd,Valderrama:2018bmv,Huang:2018wth,Lin:2019tex,Gutsche:2019eoh,Ikeno:2020vqv,Zeng:2020och,Lu:2020ste,Liu:2020yen,Ikeno:2022jpe,Hu:2022pae}. 
Actually, both pictures seem to reasonably explain the experimental data, such as mass, total decay width, and weak production from $\Omega_c$. More details on the $\Omega(2012)$ can be found in the review article of strange baryon spectrum~\cite{Hyodo:2020czb}. 

The disagreement between these two pictures comes from the predicted three-body decay. The conventional baryon should mainly decay into the two-body mode $\Xi \bar K$ rather than the three-body channel $\Xi \pi \bar K$, while the three-body final state $\Xi \pi \bar K$ are predicted to be significantly in some molecular pictures owing to the large coupling strength $|g_{\Omega(2012)\Xi^* \bar K}|$. To distinguish these two interpretations, the Belle Collaboration measure the branching fraction ratio twice as shown above. The first result favors the conventional interpretation~\cite{Belle:2019zco}, while the second one favors the molecular picture~\cite{Belle:2022mrg}. Considering the uncertainties of experiments and phenomenological models of three-body decay, we hold the opinion that one can not draw a firm conclusion of its nature until now. 

An essential fact is that the both predicted masses of the conventional $\Omega(1^2P_{3/2^-})$ state and molecular state locate in the energy region of $\Omega(2012)$. Thus, the conventional $\Omega(1^2P_{3/2^-})$ state may couples strongly with the continuum, bound states, or resonances, and the converse is also true. In Ref.~\cite{Zhong:2022cjx}, the authors calculated the coupled channel effects for $\Omega(2012)$ and found that the conventional three-quark state dominates. In the literature~\cite{Ono:1983rd,Tornqvist:1984fy,Silvestre-Brac:1991qqx,Morel:2002vk,Hwang:2004cd,Pennington:2007xr,Barnes:2007xu,Li:2009ad,Danilkin:2010cc,Zhou:2011sp,Liu:2011yp,Nagahiro:2014mba,Nagahiro:2011jn,Nagahiro:2013hba,Yang:2021tvc,Zhang:2022pxc,Ortega:2009hj,Liu:2016wxq,Ferretti:2014xqa,Garcia-Tecocoatzi:2016rcj,Lu:2017hma,Wu:2017qve,Luo:2019qkm,Luo:2021dvj,Xie:2021dwe,Ni:2021pce,Yamaguchi:2019vea}, the coupled channel effects between the bare state and hadron-hadron channels have been investigated for various systems, such as the mysterious $\sigma$, $\Lambda(1405)$, $D_{s0}^*(2317)$, $\Lambda_c(2940)$, $X(3872)$, and so on. However, many works actually considered the bare state coupled by the hadron-hadron continuum state without interactions between them. 
Several works including the hadron-hadron interactions suggested that the  interactions might be important in the coupled-channel approach~\cite{Nagahiro:2014mba,Nagahiro:2011jn,Nagahiro:2013hba,Yang:2021tvc,Zhang:2022pxc,Ortega:2009hj,Yamaguchi:2019vea}. Theoretically, the hadron-hadron interactions may dynamically generate the bound states or resonances, dynamics of which is quite different from that of the continuum states without interactions. Thus, it is crucial to study the coupled-channel effects by including the meson-baryon interactions in a unified framework. Owing to the complexity of $\Omega(2012)$, we indeed have a suitable research platform to investigate the coupled-channel effects involving meson-baryon interactions, and also it is a good opportunity to study the internal structure of this fantastic particle.

In this work, we perform a systematic analysis of the mass and strong decays for the $\Omega(2012)$ resonance in a coupled-channel approach. 
Beside considering the coupling between the bare state $\Omega(1^2P_{3/2^-})$ and $\Xi^* \bar K$ continuum states, we also include the effective $\Xi^* \bar K$ interaction via $\Omega \eta$ loop. 
Our approach has been employed in the study of the nature of $a_1(1260)$ and $\sigma$ mesons~\cite{Nagahiro:2013hba,Nagahiro:2014mba,Nagahiro:2011jn}, which figures out the bare and composite components in these two states. Here the composite states correspond to those dynamically generated by the relevant hadron-hadron interactions. 
This unified treatment can be easily extended to baryons, and suitable for investigating the nature of $\Omega(2012)$ resonance. Our results of mass, two-body decay, and three-body decay for $\Omega(2012)$ are compatible with the current experimental data, which indicates that both bare state $\Omega(1^2P_{3/2^-})$ and $\Xi^* \bar K$ channel are important and give significant contributions to the properties of $\Omega(2012)$ resonance.        

This paper is organized as follows. In Sec.~\ref{formalism}, The effective interactions and couplings are derived, and the frameworks of coupled-channel approach and strong decays are introduced as well. We present the parameters, numerical results, and discussions for the $\Omega(2012)$ resonance in Sec.~\ref{result}. A summary is given in the last section.

\section{Formalism}{\label{formalism}}

\subsection{Effective $\Xi^* \bar K$ interaction via $\Omega \eta$ loop}

\begin{figure}[!htbp]
	\centering
	\includegraphics[scale=0.68]{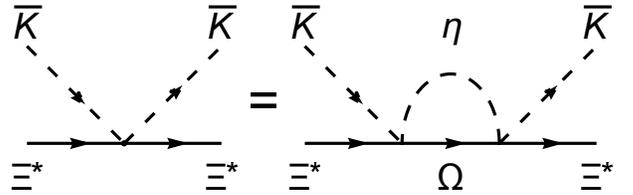}
	\vspace{0.0cm} \caption{The Feynman diagram for the effective $\Xi^* \bar K$ interaction via $\Omega \eta$ loop.}
	\label{loop}
\end{figure}

The effective $\Xi^* \bar K$ interaction is obtained by the second order process of the chiral Lagrangian going through an $\Omega \eta$ loop. The starting point is the Weinberg-Tomozawa interaction for quark-meson transitions
\begin{eqnarray}
	\mathcal{L}_{WT} = -\frac{i}{8f_p^2} \bar{q} \gamma_\mu (\phi^\mu\phi-\phi \phi^\mu) q,
	\label{L_WT_quark}
\end{eqnarray}
with
\begin{eqnarray}
\phi \equiv \vec{\lambda} \cdot \vec{M_p},  \nonumber \\
\phi^\mu \equiv \vec{\lambda} \cdot \partial^\mu \vec{M_p}.
\end{eqnarray}
Here, the $\vec{M_p}$ is the pseudoscalar meson octet field, $\vec{\lambda}$ is the SU(3) Gell-Mann matrix, and $f_p$ stands for a decay constant for a pseudoscalar meson. 
Given the leading order term, the transitions between $\Xi^* \bar K$ and $\Omega \eta$ at tree level can be written as
\begin{eqnarray}
	\mathcal {M} [\Xi^* \bar K \to \Xi^* \bar K] = \mathcal {M} [\Omega \eta \to \Omega \eta] = 0,
\end{eqnarray} 
\begin{eqnarray}
	\mathcal {M} [\Xi^* \bar K \to \Omega \eta] &=& \mathcal {M} [\Omega \eta \to \Xi^* \bar K]  \nonumber   \\&=& -\frac{3}{4f_K^2}(2\sqrt{s}-M_{\Omega}-M_{\Xi^*}),
\end{eqnarray} 
where $\sqrt{s}$ is the energy of a meson-baryon pair in their center of mass system. 
It is worth pointing that these energy dependent interactions at the quark level leads to the interaction for hadrons~\cite{Pavao:2018xub,Ikeno:2020vqv} when taking the baryon matrix elements of 
(\ref{L_WT_quark}) by the standard quark model wave functions.
The direct interaction of $\Xi^* \bar K$ channel itself vanishes. 
Hence, the effective potential $v_{\rm com}$ for $\Xi^* \bar K$ channel can be obtained through the $\Omega \eta$ loop, which is shown in Figure~\ref{loop}. After the nonrelativistic reduction, the integrated $\Omega \eta$ two-body propagator can be expressed as
\begin{eqnarray}
	G_{\Omega \eta}(\sqrt{s}) &=& \frac{M_\Omega}{2\pi^2} \int_0^\Lambda d|\boldsymbol{q}| \frac{\boldsymbol{q}^2}{s-(E_\Omega+\omega_\eta)^2+i \epsilon}   \nonumber   \\ &&\times \frac{E_\Omega+\omega_\eta}{E_\Omega \omega_\eta}\Bigg(1+\frac{2\boldsymbol{q}^2}{9 M_\Omega}\Bigg),
	\label{G_Omega_eta}
\end{eqnarray} 
where the $E_\Omega=\sqrt{M_\Omega^2+\boldsymbol{q}^2}$ and $\omega_\eta=\sqrt{m_\eta^2+\boldsymbol{q}^2}$ are the energies for intermediate $\Omega$ and $\eta$ hadrons, respectively. The $\Lambda$ is the cut off parameter to eliminate the ultraviolet divergence, which reflects the finite size of baryons. The term $\Bigg(1+\frac{2\boldsymbol{q}^2}{9 M_\Omega}\Bigg)$ arises from the propagator of spin-3/2 particle, which gives minor effects for small momentum $|\boldsymbol{q}|$. 
The resulting effective potential for  $\Xi^* \bar K$  composite states is then given by
\begin{eqnarray}
	v_{\rm com}= \mathcal {M} [\Xi^* \bar K \to \Omega \eta] G_{\Omega \eta}(\sqrt{s}) \mathcal {M} [\Omega \eta \to \Xi^* \bar K].
\end{eqnarray}   

\subsection{Interaction between the bare state and $\Xi^* \bar K$ channel}

To calculate the mixture between the bare state $\Omega(1^2P_{3/2^-})$ and $\Xi^* \bar K$ channel, one should extract the effective interaction from quark-meson couplings. Here, we adopt the axial-vector type coupling for the interaction between the pseudoscalar meson and a light quark inside the baryons. It can be written as
\begin{eqnarray}
	\mathcal{L}_{M_p qq} = \frac{g_A^q}{2f_p} \bar{q} \gamma_\mu\gamma_5 \phi^\mu q,
	\label{L_Mqq}
\end{eqnarray}
where $g_A^q$ is the quark-axial-vector coupling strength. This interaction is inspired by the low-energy theorem of chiral symmetry and has been extensively employed in the quark model calculations~\cite{Xiao:2018pwe,Liu:2019wdr,Arifi:2022ntc,Nagahiro:2016nsx}. 
We note that the interaction (\ref{L_Mqq})  when $g_A^q = 1$ is found in the non-linear chiral meson-quark Lagrangian consistently with the Weinberg-Tomozawa type interaction (\ref{L_WT_quark}).  
To match the nonrelativistic wave functions of conventional baryons, the interaction operator should be expanded in powers of $1/m$, and the terms up to order $1/m$ is enough for the present low momentum phenomena. The nonrelativistic form can be expressed as
\begin{eqnarray}
	\mathcal{H}_{\rm NR} = \frac{g_A^q}{2f_p}  \left[ \boldsymbol{\sigma}\cdot \boldsymbol{q} + \frac{\omega}{2m}\left( \boldsymbol{\sigma}\cdot \boldsymbol{q} -2 \boldsymbol{\sigma}\cdot \boldsymbol{p}_i \right) \right]. \label{nonrel}
\end{eqnarray}
Here, the $m$ is the constituent quark mass, $q=(\omega, \boldsymbol{q})$ corresponds to the 4-momentum of the outgoing pseudoscalar meson, and $\boldsymbol{p}_i$ stand for the momentum of a quark inside the initial baryon.

The wave functions of the $\Omega(1^2P_{3/2^-})$ and $\Xi^*$ resonances can be easily constructed 
in the standard quark model.  
For the spatial part, the harmonic oscillator wave functions are adopted here. Explicit formulas of these wave functions can be found in Ref.~\cite{Nagahiro:2016nsx}. Then, we can calculate the amplitude $\Omega(1^2P_{3/2^-}) \to \Xi^* \bar K$ straightforwardly
\begin{eqnarray}
\mathcal {M} [\Omega(1^2P_{3/2^-}) \to \Xi^* \bar K] 
&=& \frac{g_A^q \omega_{\bar K} \alpha}{f_K m} e^{-\frac{\boldsymbol{q}^2}{9\alpha^2}},  
\end{eqnarray}
where the constant $\alpha$ is the inverse range parameter of the harmonic oscillator wave functions. 
The Gaussian functional form corresponds to a form factor, which plays the role of making the relevant loop
$\Xi^* \bar K$ integral finite.  
In practice, as shown in (\ref{G_Xi_K_2}), the Gaussian form factor is replaced by the sharp cutoff $\Lambda$ in momentum space, 
as consistently done also for the $\Omega\eta$ loop function (\ref{G_Omega_eta}).  

\begin{figure}[!htbp]
	\centering
	\includegraphics[scale=0.47]{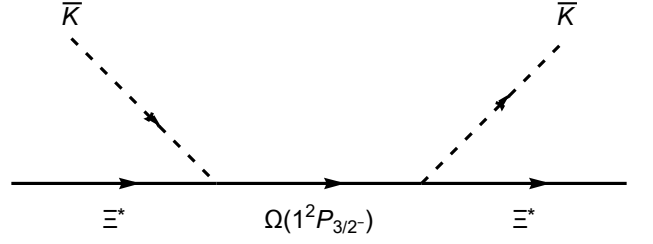}
	\vspace{0.0cm} \caption{The tree-level Feynman diagram for the interaction between the bare state and $\Xi^* \bar K$ channel.}
	\label{tree}
\end{figure}

Finally, the tree diagram $\Xi^* \bar K \to \Omega(1^2P_{3/2^-}) \to \Xi^* \bar K$ as shown in Figure~\ref{tree} for the bare state induced potential $v_{\rm bare}$ can be expressed as
\begin{eqnarray}
	v_{\rm bare} &=& \frac{2M_0g_A^{q2} \alpha^2(\sqrt{s}-M_{\Xi^*})^2}{f_K^2 m^2(s-M_0^2)},
\label{coupling_bareOmega_XiK}	
\end{eqnarray}
where the mass $M_0$ is for the bare state $\Omega(1^2P_{3/2^-})$. 
Thus, the full interaction $v_{\rm full}$ is
\begin{eqnarray}
	v_{\rm full}=v_{\rm com}+v_{\rm bare}.  
\end{eqnarray} 

\subsection{The two-level problem}

With the interactions $v_{\rm com}$, $v_{\rm bare}$, and $v_{\rm full}$, we can obtain the amplitudes $t_{\rm com}$ and $t_{\rm full}$ by using the Bethe-Salpeter equation
\begin{eqnarray}
	t_{\rm com}=\frac{v_{\rm com}}{1-v_{\rm com}G_{\Xi^* \bar K}},
\end{eqnarray}  
\begin{eqnarray}
	t_{\rm full}=\frac{v_{\rm full}}{1-v_{\rm full}G_{\Xi^* \bar K}}=\frac{v_{\rm com}+v_{\rm bare}}{1-(v_{\rm com}+v_{\rm bare})G_{\Xi^* \bar K}}.
\end{eqnarray}
The loop function $G_{\Xi^* \bar K}$ has the similar structure to $G_{\Omega \eta}$, and can be written as
\begin{eqnarray}
	G_{\Xi^* \bar K}(\sqrt{s}) &=& \frac{M_{\Xi^*}}{2\pi^2} \int_0^\Lambda d|\boldsymbol{q}| \frac{\boldsymbol{q}^2}{s-(E_{\Xi^*}+\omega_{\bar K})^2+i \epsilon}   \nonumber   \\ &&\times \frac{E_{\Xi^*}+\omega_{\bar K}}{E_{\Xi^*} \omega_{\bar K}}\Bigg(1+\frac{2\boldsymbol{q}^2}{9 M_{\Xi^*}^2}\Bigg).
	\label{G_Xi_K_2}
\end{eqnarray}  
Then, we can find the solution of the full amplitude in these composite and bare bases, and compare the obtained poles with the physical state $\Omega(2012)$. This problem is quite similar to the two-level problem in quantum mechanics, and the idea has been employed to the study of the nature of $a_1(1260)$ and $\sigma$ mesons~\cite{Nagahiro:2013hba,Nagahiro:2014mba,Nagahiro:2011jn}.  Here, we briefly introduce this approach.

Let us first express the amplitude $t_{\rm com}$ for $\Xi^*\bar K$ scatterings by an $s-$channel pole term as
\begin{equation}
	t_{\rm com}
	\equiv g_R(s)\frac{1}{\sqrt{s}-\sqrt{s_p}}g_R(s) \ ,
\end{equation}
where $\sqrt{s_p}$ is the pole position of the amplitude {$t_{\rm com}$}, and $g_R(s)$ is a function corresponding to 
the coupling strength of the obtained composite particle at $s = s_p$. 
Next we see that the relation between $v_{\rm bare}$ and coupling strength $g(s)$ of bare $\Omega(1^2P_{3/2^-})$ state can be defined as
 \begin{equation}
 	v_{\rm bare}
 	\equiv g(s)\frac{1}{\sqrt{s}-M_0}g(s)\ .
 \end{equation}
Then, we can express the full scattering amplitude by
\begin{equation}
	t_{\rm full}=(g_R,g)
	\frac{1}{\hat{D}_0^{-1}-\hat{\Sigma}}
	\begin{pmatrix}
		g_R  \\ g
	\end{pmatrix} \ ,
	\label{eq:T_fullM}
\end{equation}
with 
\begin{equation}
	\hat{D}_0^{-1} = 
	\begin{pmatrix} \sqrt{s}-\sqrt{s_p} & 0  \\ 0  & \sqrt{s}-M_0 \end{pmatrix}, 
\end{equation}
\begin{equation}
	\hat{\Sigma} = 
	\begin{pmatrix}
		0 & g_R G_{\Xi^* \bar K} g \\ 
		g G_{\Xi^* \bar K} g_R & g G_{\Xi^* \bar K} g
	\end{pmatrix} \ .
\end{equation}
The diagonal elements of $\hat{D}_0$ are the free propagators of two
different states, one for the composite state and the other for
the bare state.  The matrix $\hat{\Sigma}$ represents the self-energy and mixing interaction between these two states. The detailed derivation can be found in Ref.~\cite{Nagahiro:2013hba}, and a slight modification here arises from the  different propagators between baryons and mesons.

Now, the matrix
\begin{equation}
	\hat{D} = \frac{1}{\hat{D}_0^{-1}-\hat{\Sigma}}
	\label{eq:full_D}
\end{equation}
is the full propagators of the physical states represented by the two bases of the composite and bare ones. We adopt the $M^*$ to denote the poles obtained by the full $t$-matrix $t_{\rm full}$, which correspond to the physical states and can be compared with the experimental observations. 
The residues of the
diagonal elements $D^{ii}$ are obtained by
\begin{equation}
	z^{ii}=\frac{1}{2\pi i} \oint_\gamma D^{ii}(s)\ d\sqrt{s}\ ,  \ \ (i=1,2)
\end{equation}
where $\gamma$ are the closed circles around the poles 
$M^*$.  
The residues $z^{11}$ and $z^{22}$ reflect the information of the corresponding components. 
For instance, $D^{22}$ is the full propagator of the bare state, and its residue $z^{22}$ has the meaning of the probability of finding the bare component in the physical state. 
Thus, we can schematically express the physical state $\Omega(2012)$ as
\begin{equation}
\Omega(2012) = \sqrt{z^{11}}|1\rangle +
	\sqrt{z^{22}}|2\rangle,
\end{equation}
where the $|1\rangle$ and $|2\rangle$ are the composite $\Xi^* \bar K$ and bare $\Omega(1^2P_{3/2^-})$ states, respectively.

\subsection{Strong decays}

The physical state $\Omega(2012)$ can decay into the $\Xi \bar K$ and $\Xi \pi \bar K$ channels. For the two-body decay, we can adopt the transition operator $\mathcal{H}_{\rm NR}$ together with the wave functions of baryons to estimate the helicity amplitude $\mathcal{A}_h$~\cite{Nagahiro:2016nsx}. 
Then, the width of the two-body decay $\Omega(2012) \to \Xi \bar K$  can be written as
\begin{eqnarray}
	\Gamma[{\Omega(2012) \to \Xi \bar K}]
	&=& z^{22} \frac{1}{4\pi} 
	\frac{|\boldsymbol{q}|}{2M_i^2}\frac{1}{2J+1}\sum_h |\mathcal{A}_h|^2,
	\label{Gamma2-body}
\end{eqnarray} 
where $|\boldsymbol{q}|$ is the momentum of outgoing meson, $M_i$ and $J$ are the mass and spin of the initial baryon, respectively. 
We emphasize that in our model, the two-body $\Xi \bar K$ decay arises from its coupling to the bare  component.  
Because the coupling is determined by (\ref{L_Mqq}), there is no free parameter.  
In contrast, in the pure molecular (composite) picture, there is no coupling for 
the transition $\Xi^* \bar K \to \Xi \bar K$. 
Therefore, for instance, in Refs.~\cite{Pavao:2018xub,Ikeno:2020vqv}, the authors introduced two parameters 
to describe this process by assuming that it gives a dominating contribution to the total decay width.  

\begin{figure}[!htbp]
	\centering
	\includegraphics[scale=0.47]{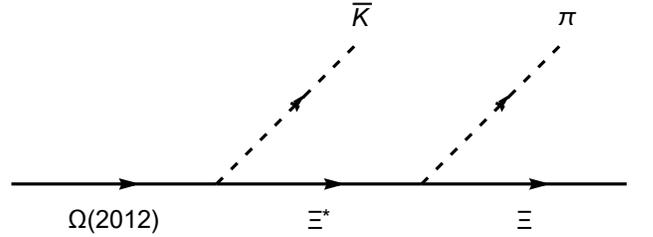}
	\vspace{0.0cm} \caption{The diagram of sequential decay chain $\Omega(2012) \to \Xi^* \bar K \to \Xi \pi \bar K$.}
	\label{decay1}
\end{figure}

To calculate the partial decay width of channel $\Xi \pi \bar K$, we employ the same formula as Ref.~\cite{Ikeno:2020vqv}. 
The sequential decay chain $\Omega(2012) \to \Xi^* \bar K \to \Xi \pi \bar K$ is assumed for this process, the diagram of which is shown in Figure~\ref{decay1}. 
We can obtain the coupling strengthen between the physical state $\Omega(2012)$ and $\Xi^* \bar K$ by
\begin{equation}
	g_{\Omega(2012)\Xi^* \bar K}^2 =\lim_{\sqrt{s} \to M^*}(\sqrt{s} - M^*) t_{\rm full}=\frac{1}{2\pi i} \oint_\gamma  t_{\rm full}(s) d\sqrt{s}.
	\label{g2_OmegaXi*K}
\end{equation}
Then, we take into account the phase space factor for the $\Xi^*$ decay width in the convolution integral as
\begin{equation}
	\Gamma_{\Xi^*}=\Gamma_{\Xi^*, \rm{on}} \frac{ | \tilde {\boldsymbol q}_\pi|^3}{ | \tilde {\boldsymbol q}_{\pi, {\rm on}}|^3} \ \theta(M_{\rm inv}(\pi \Xi) - m_\pi -M_\Xi),
\end{equation}
where $ | \tilde {\boldsymbol q}_\pi|$, $| \tilde {\boldsymbol q}_{\pi, {\rm on}}|$ are the momentum with masses of $M_{\rm inv} (\pi \Xi)$ and $M_{\Xi^*}$ for  the $\Xi^*$ baryon. Also, the other choice of $\Gamma_{\Xi^*}$ is the constant width from the experimental data, and we will discuss the differences between these two ways in Sec.~\ref{result}. Finally, we can calculate the three-body decay straightforwardly
\begin{eqnarray}
	&&\frac{d \Gamma[\Omega(2012) \to \Xi^* \bar K \to \Xi \pi \bar K]}{ d M_{\rm inv}(\pi \Xi)} \nonumber  \\ &=& \frac{1}{(2\pi)^2} \frac{M_{\Xi^*}}{M_{\Omega(2012)}}  | \tilde {\boldsymbol q}_{\bar K}|  g_{\Omega(2012)\Xi^* \bar K}^2
	\frac{\Gamma_{\Xi^*}}{\left| M_{\rm inv}(\pi \Xi) -M_{\Xi^*} +i \frac{\Gamma_{\Xi^*}}{2}\right|^2}.\nonumber  \\ 
\end{eqnarray} 

\section{Results and discussions}\label{result}

\subsection{Parameters}

To investigate the mass and decay behaviors of the $\Omega(2012)$, several parameters should be fixed. For the composite case, we need the decay constant $f_K$ and cut off $\Lambda$. In Ref.~\cite{Ikeno:2020vqv}, the $f_\pi=93$ MeV and $\Lambda=700$ MeV are adopted to generate a pole slightly below the $\Xi^* \bar K$ threshold. If we use the same parameters and neglect the $\Xi \bar K$ decay channel, we can also obtain the similar loosely bound state that corresponds to the physical state $\Omega(2012)$. This is consistent with the molecular interpretations at the hadron level. Since only strange mesons and multistrangeness baryons are considered in this work, the choice $f_p=f_K=111~\rm{MeV}$ is more suitable in the Lagrangians~(\ref{L_WT_quark}) and~(\ref{L_Mqq})~\cite{Arifi:2022ntc}. Also, we employ the $\Lambda=500$ MeV here which we consider is a reasonable choice for conventional baryons.

The quark axial-vector coupling constant we employ $g_A^q=1$.
In some literatures~\cite{Xiao:2013xi,Yamaguchi:2019vea,Nagahiro:2016nsx}, a smaller value was used to effectively reproduce the decays of $K^* \to K\pi,\ D^* \to D \pi$, $\Sigma_c \to \Lambda_c \pi$ and $\Xi^* \to \Xi \pi$. The reduction of $g_A^q$ from unity is also needed to explain the nucleon's $g_A$ in the quark model (there are many references for this; see for instance~\cite{Hosaka:1996ee}).  
One possible reason is the relativistic effect of the quark motion in a hadron. All of these cases are transitions of the hadrons with quarks in the orbitally ground state. However, it was demonstrated in Ref.~\cite{Arifi:2021orx,Arifi:2022ntc} the correction is not important for transitions between a $P$-wave excited state to the ground state, which is the case that we need.  
Lastly, for the interactions in the $\Omega(1^2P_{3/2^-}) \to \Xi^{(*)} \bar K$ processes, we adopt the average quark mass $m=450$ GeV  and harmonic oscillator parameter $\alpha = 331$ MeV following the previous work~\cite{Arifi:2022ntc}.   

In the literature, there are lots of calculations for the mass spectrum of $\Omega$ family, and the estimated mass for $\Omega(1^2P_{3/2^-})$ state varies from to 1953 to 2142 MeV~\cite{Oh:2007cr,Capstick:1986bm,Chao:1980em,Faustov:2015eba,Chen:2009de,Bijker:2015gyk,Pervin:2007wa,Engel:2013ig,Liu:2019wdr}. 
These masses correspond to the physical ones after the renormalization due to the meson-baryon loops. 
In the unquenched approach, the bare states are supposed to be heavier than the physical state, and the effects of coupled channels or hadron loops reduce the masses such that the bare states are pushed down to those of physical states~\cite{Pennington:2007xr,Barnes:2007xu,Li:2009ad,Ferretti:2014xqa,Garcia-Tecocoatzi:2016rcj,Luo:2019qkm,Liu:2016wxq,Lu:2017hma,Wu:2017qve}. 
These mass shifts are predicted to be from tens to hundreds MeV, depending on the physical systems and phenomenological models. 
Hence, we should choose a larger mass $M_0$ for the bare state relative to the theoretical predictions and experimental observation of $\Omega(2012)$. 
In the present study, we adopt $M_0=2150$ MeV for the bare state $\Omega(1^2P_{3/2^-})$, 
and discuss the influence of other choices. 

\subsection{Poles}

By analyzing the amplitude $t_{\rm com}$, we find a pole of composite state in the second Riemann sheet (SRS) 
above the $\Xi^* \bar K$ threshold.
In doing so, we have defined the analytic structure of the complex loop function
\begin{equation}
	G^{\text{SRS}}_i (\sqrt{s}) = G_i(\sqrt{s}) + \left\{\begin{matrix}
		0, &  \text{for } \text{Re}(\sqrt{s})<\sqrt{s}_{th}\\ 
		i\frac{M_i |\boldsymbol{q}|}{2 \pi \sqrt{s}}\Bigg(1+\frac{2\boldsymbol{q}^2}{9 M_i^2}\Bigg), &  \text{for } \text{Re}(\sqrt{s})>\sqrt{s}_{th}
	\end{matrix}\right. ,
\end{equation}
with
\begin{equation}
	|\boldsymbol{q}| = \frac{\sqrt{[s-(M_i+m_i)^2][s-(M_i-m_i)^2]}}{2 \sqrt{s}},
\end{equation}
where the $i$ stands for $\Xi^* \bar K$ or $\Omega \eta$ channel. 
The pole locates at $2139.2 - 124.3$ MeV, which represents a broad resonance generated by the effective $\Xi^* \bar K$ interaction $v_{\rm com}$.
Obviously, such a state cannot be identified with the observed  $\Omega(2012)$. 

So, now we couple the bare state $\Omega(1^2P_{3/2^-})$ to the broad composite state by including the interaction (\ref{coupling_bareOmega_XiK}).  
With the amplitude $t_{\rm full}$, we obtain two poles: one is located at $2007.9$ MeV, and the other is at $2236.8-69.3i$ MeV. 
The pole at $2007.9$ MeV is below the $\Xi^* \bar K$ threshold, and exists as a loosely bound state. This mass is consistent with the experimental data $2012.5 \pm 0.7 \pm 0.5~\rm{MeV}$, which would be a good candidate of $\Omega(2012)$. 
Certainly, without considering the decay channel $\Xi \bar K$ in full amplitude $t_{\rm full}$, we obtain a loosely bound state rather than a resonance. The small decay width of $\Omega(2012)$ can be calculated  perturbatively by introducing the transition operators in the following subsection.  

The pole at $2236.8-69.3i$ MeV is far away from $\Omega(2012)$. 
It is tempting to identify it with other observed excited $\Omega$ state.  
In the Review of Particle Physics~\cite{ParticleDataGroup:2022pth}, a structure $\Omega(2250)$ exists, which has a mass of $2252\pm9$ MeV and a width of $55\pm18$ MeV. 
Also, this resonance decays into $\Xi^* \bar K$ and $\Xi \pi \bar K$ channels experimentally.
In the previous theoretical studies~\cite{Xiao:2018pwe,Liu:2019wdr},  a possible interpretation as $\Omega(1D)$ was given. 
However, due to lack of detailed experimental information, we cannot determine the nature of the observed states $\Omega(2250)$.

Moreover, we can obtain the composite and bare components from the $t_{\rm full}$. 
Focusing on the pole at $2007.9$ MeV, we find that the probability of bare component is $28.7\%$. 
This result suggests that both components of the bare state and $\Xi^* \bar K$ channel are important, and the coupled-channel effect plays an essential role for the better understanding of the nature of $\Omega(2012)$.

\subsection{Decays}

With the formula (\ref{Gamma2-body}) given in Sec.~\ref{formalism}, it is straightforward to evaluate the two-body decay width. 
We have found that 
\begin{eqnarray}
	\Gamma[\Omega(2012) \to \Xi \bar K]  = 3.02~\rm{MeV}.
\end{eqnarray} 
In comparison with the total decay width, this result is consistent with the experimental measurement.
Also, it should be emphasized that in our model the two-body decay width arises from the three-quark component in the physical $\Omega(2012)$ state without introducing additional parameters.  

\begin{figure}[!htbp]
	\centering
	\includegraphics[scale=0.47]{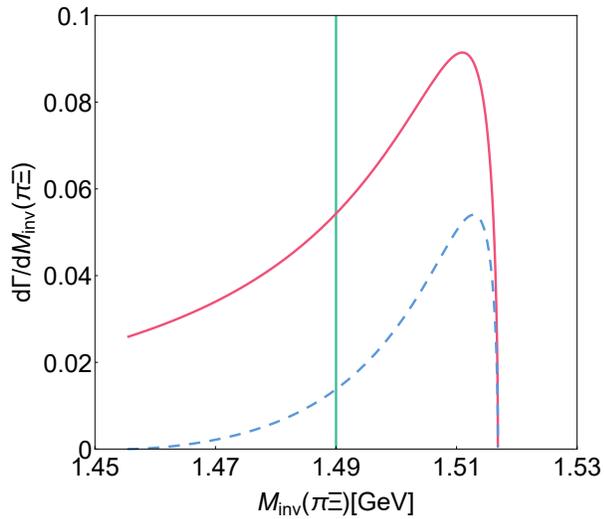}
	\vspace{0.0cm} \caption{The invariant mass distribution $d \Gamma[\Omega(2012) \to \Xi^* \bar K \to \Xi \pi \bar K]/ d M_{\rm inv}(\pi \Xi)$ for three-body channel. The red solid and blue dashed lines correspond to energy independent and dependent $\Xi^*$ widths, respectively. The green vertical line is the lower cut 1.49 GeV as implemented by Belle experiment in 2019~\cite{Belle:2019zco}.}
	\label{decay2}
\end{figure}

Now for three-body decays, we consider the sequential process $\Omega(2012) \to \Xi^* \bar K \to \Xi \pi \bar K$.  
For this, we need the coupling strength $|g_{\Omega(2012)\Xi^* \bar K}|$.  
Using (\ref{g2_OmegaXi*K}), 
the coupling 
near the physical pole is predicted to be 1.99, which is similar to the result of chiral unitary approach~\cite{Ikeno:2020vqv}. 
This indicates that we should get the analogous results to that of molecular picture. 
We plot the invariant mass distribution $d \Gamma[\Omega(2012) \to \Xi^* \bar K \to \Xi \pi \bar K]/ d M_{\rm inv}(\pi \Xi)$ in Figure~\ref{decay2}, where both energy independent and dependent $\Xi^*$ widths are considered as in Ref.~\cite{Ikeno:2020vqv}. 
After the integration over $M_{\rm inv}(\pi \Xi)$, the three-body decay widths can be obtained. Given the strong decays, the total decay width of $\Omega(2012)$ equals to the sum of partial widths for two-body and three-body channels. 
Also, we can estimate the theoretical ratio of the branching fractions $\mathcal {R}_{\Xi \bar K}^{\Xi \pi \bar K}$. 
The decay properties for $\Omega(2012)$ under various situations are listed in Table~\ref{table}. 
It should be noted that the lower cut 1.49 GeV in $M_{\rm inv}(\pi \Xi)$ was implemented by Belle Collaboration in 2019~\cite{Belle:2019zco}, which should be included for the comparisons of ratio of branching fractions.

\begin{table}[!htbp]
	\begin{center}
		\caption{\label{table} Theoretical results of the strong decays for the $\Omega(2012)$ resonance in MeV. The lower cut affects the three-body decays, ratio of branching fractions, and total width.}
		\renewcommand{\arraystretch}{1.5}
		\begin{tabular*}{8.5cm}{@{\extracolsep{\fill}}p{1.6cm}<{\centering}p{1.2cm}<{\centering}p{1.2cm}<{\centering}p{1.2cm}<{\centering}p{1.2cm}<{\centering}}
			\hline\hline
			
			\multirow{2}{*}{Decays}  &  \multicolumn{2}{c}{No cut} &  \multicolumn{2}{c}{With cut} \\
			\cline{2-3}    \cline{4-5}    
			& Independent & Dependent & Independent & Dependent  \\ \hline
			$\Xi \bar K$ & \multicolumn{4}{c}{$3.02$}  \\
            $\Xi \pi \bar K$ & 3.27  & 1.07 & 1.99  & 0.92 \\
			$\Gamma[\Omega(2012)]$ & 6.29  & 4.09 & 5.01  & 3.94 \\
			 $\mathcal {R}_{\Xi \bar K}^{\Xi \pi \bar K}$         & $1.08$ &  $0.35$  & $ 0.66$ &  $ 0.30$   \\\hline
			  $\Gamma[\Omega(2012)]_{exp}$ &  \multicolumn{4}{c}{$6.4^{+2.5}_{-2.0} \pm 1.6$~\cite{Belle:2018mqs}} \\
			   ${\mathcal {R}_{\Xi \bar K}^{\Xi \pi \bar K}}_{exp}$    &    \multicolumn{2}{c}{ $0.97 \pm 0.24 \pm 0.07$~\cite{Belle:2022mrg}} & \multicolumn{2}{c}{$\textless~0.119$~\cite{Belle:2019zco}}  \\
			\hline\hline
		\end{tabular*}
	\end{center}
\end{table}

It can be seen that the energy independent and dependent $\Xi^*$ widths actually affect the three-body decay width significantly. 
With the energy independent $\Xi^*$ width and no lower cut case, the calculated total decay width and ratio of branching fractions are
\begin{eqnarray}
	\Gamma[\Omega(2012)] = 6.29,
\end{eqnarray} 
\begin{eqnarray}
\mathcal {R}_{\Xi \bar K}^{\Xi \pi \bar K} = 1.08,
\end{eqnarray} 
which agree well with the observed total decay width~\cite{Belle:2018mqs} and new experimental measurement $\mathcal {R}_{\Xi \bar K}^{\Xi \pi \bar K}$ in 2022~\cite{Belle:2022mrg}. Moreover, for other three cases, the calculated ratio of branching fractions $\mathcal {R}_{\Xi \bar K}^{\Xi \pi \bar K}$ disagree with the experimental data.

Although the present calculations with energy independent $\Xi^*$ width give good descriptions of new experimental data, 
the energy dependent $\Xi^*$ width should be used for the formula of three-body decay theoretically. 
We find that the three-body calculations are sensitive to the phenomenological models.  Also, the two experimental measurements about $\mathcal {R}_{\Xi \bar K}^{\Xi \pi \bar K}$ by Belle Collaboration are in conflict with each other. 
It is hoped that more theoretical and experimental efforts is needed to investigate the decay behaviors of $\Omega(2012)$ resonance.

\subsection{Further discussions}

Our present choice of bare mass for $\Omega(1^2P_{3/2^-})$ is $2150$ MeV, which is slightly higher than the pure composite pole at $2139.2 - 124.3i$ MeV. Also, one may reduce the quark axial-vector coupling constant $g_A^q$ as in Refs.~\cite{Xiao:2013xi,Yamaguchi:2019vea,Nagahiro:2016nsx,Hosaka:1996ee}. It is interesting to investigate the variations of these parameters. An alternative choice with $M_0=2070$ MeV and $g_A^q=0.6$ can also describe the properties of $\Omega(2012)$. In this case, the bare mass is smaller than the composite pole, which reverses the mass order of the two-level problem. The generating pole is at 2013.9 MeV with the three-quark component of $46.7\%$, and the decay behaviors also roughly agree with the Belle experiments. Moreover, the cutoff parameter $\Lambda$ can be varied around 500 MeV. It is found that when we increase 
$\Lambda$, the energy of the pole near $\Xi^* \bar K$ threshold decreases, and vice versa. Certainly, these changes of pole positions induced by $\Lambda$ can be adjusted by the parameter $M_0$ within reasonable range.

In the present calculation, we neglect the coupling between the bare state $\Omega(1^2P_{3/2^-})$ and baryon-meson channel $\Omega \eta$. Indeed, this coupling can cause some corrections to the full amplitude $t_{\rm full}$ and positions of poles indirectly. After including this coupling, we found that the pole position at 2007.9 MeV change only slightly, which does not affect our results and conclusions in present work. For the second pole near $\Omega \eta$, this coupling may affect significantly, and it is expected to study this energy region in future.

In theory, we can discuss the pole flow by varying the coupling strength between the bare state and $\Xi^* \bar K$ like the $\sigma$ meson case. For this energy dependent interaction, understanding the the pole flow and compositeness is a very intriguing topic~\cite{Nagahiro:2014mba}. By changing the bare mass, cut off, and coupling strength, we can obtain various playgrounds to investigate the specific behaviors of poles. This is beyond the scope of the current study of $\Omega(2012)$, and we prefer to examine them in our future work.

Our present results of masses and decay behaviors are consistent with current experimental observations of $\Omega(2012)$, which suggests that both three-quark and $\Xi^* \bar K$ components are significant and important. Indeed, without the three-quark component, the pole generated by effective $\Xi^* \bar K$ interaction is far away from the $\Omega(2012)$, and no leading order term contributes to the two-body decay channel $\Xi \bar K$. Still, without the $\Xi^* \bar K$ component, the pure three-quark core will give small three-body partial decay width, which is not favored by new experimental measurement $\mathcal {R}_{\Xi \bar K}^{\Xi \pi \bar K}$.   

Interestingly, various interpretations in the literature, such as the conventional $\Omega(1P)$ baryon, molecular picture, and mixture dominated by bare state, seems to reasonably or partly describe the experimental data with several model-dependent parameters. In fact, these explanations together with ours give different internal structures of $\Omega(2012)$, but current experiments are insufficient to distinguish them irrefutably. More experimental information are needed, and further theoretical investigations on other analogous systems may be also helpful to disentangle this puzzle.   

\section{SUMMARY}
 
In this work, we study the mass and strong decay behaviors of the $\Omega(2012)$ resonance in a coupled-channel approach. The mixture of bare $\Omega(1^2P_{3/2^-})$ state and $\Xi^* \bar K$ channel are considered. Also, the meson-baryon interactions are included, which manifests as the effective $\Xi^* \bar K$ interaction. With these interactions, we obtain a pole that has a mass of $2007.9$ MeV by solving the Bethe-Salpeter equation. The strong decay behaviors are also investigated, and the predicted total decay widths and ratio of branching fractions $\mathcal {R}_{\Xi \bar K}^{\Xi \pi \bar K}$ agree well with the new experimental measurement by Belle Collaboration. Our results indicate that both three-quark core and  $\Xi^* \bar K$ channel are important and play essential roles in describing the mass and strong decay behaviors of $\Omega(2012)$ resonance.

Despite these successes, we notice that the calculations of three-body decay are sensitive to the phenomenological models.  
Because of this sensitivity, the previous interpretations in the literature can reasonably or partly describe the experimental data. 
However, given the uncertainties of the experimental data, it is still early to draw a final conclusion for the structure of the $\Omega(2012)$ resonance.  
Besides the mass and strong decay behaviors, the weak and electromagnetic properties may be helpful for us to differentiate various interpretations. 
More experimental and theoretical efforts on $\Omega(2012)$ resonance and other analogous systems are needed to clarify its nature.

\bigskip
\noindent
\begin{center}
	
	{\bf ACKNOWLEDGEMENTS}\\
	
\end{center}
We would like to thank Ahmad Jafar Arifi for helpful discussions. This work is supported by the National Natural Science Foundation of China under Grants No. 11705056 and No. U1832173, the Key Project of Hunan Provincial Education Department under Grant No. 21A0039, and the State Scholarship Fund of China Scholarship Council under Grant No. 202006725011. A.H.  is  supported by the Grants-in-Aid for Scientific Research (Grant Numbers 21H04478(A)) and the one on Innovative Areas (No. 18H05407), and H.N. by the Grants-in-Aid for Scientific Research (Grant No. JP21K03536 (C)).

\end{document}